\documentstyle[12pt]{article}
\begin{document}
\title{COMMENT ON PLAGA'S ANALYSIS OF NEUTRINO SCATTERING CHARM II
 DATA}
\author{D. PALLE \\Department of Theoretical Physics,Rugjer Bo\v skovi\'c 
Institute, \\P.O.Box 1016,Zagreb,CROATIA}
\date{ }
\maketitle

Recently Plaga \cite{Plaga} claimed that the existence of a vector
coupling of neutrinos in neutral currents, observed in the Charm II 
collaboration \cite{Charm}, led to the conclusion that neutrinos
were Dirac particles. Although he successfully defended \cite{Plaga,Kayser}
some of his assertions, his claim that Charm II data inevitably
required Dirac neutrinos is not necessarily correct.

Namely, Charm II data are fitted with the Standard Model, thus
supposing no flavour mixing of leptons. Solar and  atmospheric-
neutrino, as well as LSND neutrino \cite{LSND} measurements suggest
the existence of neutrino flavour mixing. One should fit the
Charm II data including also the charged-current contribution in the
electron- $\mu$-neutrino scattering. Following 't Hooft \cite{tHooft},
one can immediately find the effective couplings for Dirac neutrinos:

\begin{eqnarray}
(\nu_{e},e): g_{V}=\mid U_{ee}\mid ^{2}-\frac{1}{2}+2 sin^{2}\theta_
{W}\ ,\ g_{A}=\mid U_{ee}\mid ^{2}-\frac{1}{2}, \nonumber \\
(\nu_{\mu},e): g_{V}=\mid U_{e\mu}\mid ^{2}-\frac{1}{2}+
2 sin^{2}\theta_{W}\ ,\ g_{A}=\mid U_{e\mu}\mid ^{2}-\frac{1}{2}.
 \nonumber
\end{eqnarray}

Similarly, one can study the scattering with Majorana neutrinos.
Neglecting the mass terms, the cross sections of the (e,$\mu$)-
neutrino-electron scattering \cite{tHooft} 

\begin{eqnarray}
\frac{d\sigma}{d E_{e}} \propto E^{2}_{\nu}(g_{V}+g_{A})^{2}+
(E_{\nu}-E_{e})^{2}(g_{V}-g_{A})^{2}\  \nonumber
\end{eqnarray}

should be used to fit the Charm II data. Since $\frac{1}{2}-
2 sin^{2}\theta_{W}\simeq 0.04$, the impact of the 
neutrino mixing with the same magnitude $\mid U_{e\mu}\mid ^{2}=
{\cal O}(10^{-2})$ to the fit of the effective couplings
could be considerable. Notice that the first results of NOMAD\cite{NOMAD} 
do not exclude smaller neutrino masses (less than 1 eV)
and larger mixing in the LSND plot
 \cite{LSND}.

To conclude, one can say that it is premature to draw a conclusion 
concerning the Dirac or Majorana nature of the neutrino until  
there is no electroweak theory (that can explain small neutrino
masses and flavour mixing in a natural way) to fit all the 
existing data.

\end{document}